\documentclass[12pt, a4paper]{elsarticle}

\usepackage{amssymb}

\usepackage{amsmath}

\usepackage{xcolor}

\usepackage{float}
\restylefloat{table}

\journal{SoftwareX}

\begin{document}

\begin{frontmatter}

\title{Numerical package for solving the JIMWLK evolution equation in C++}

\author[label1,label2]{Piotr Korcyl}

\address[label1]{Institute of Theoretical Physics, Jagiellonian University, ul. \L ojasiewicza 11, 30-348 Krak\'ow, Poland}
\address[label2]{Institut f\"ur Theoretische Physik, Universit\"at Regensburg, D-93040 Regensburg, Germany}

\begin{abstract}
Precise and detailed knowledge of the internal structure of hadrons is one of the most actual problems in elementary particle physics. In view of the planned high energy physics facilities, in particular the Electron-Ion Collider constructed in Brookhaven National Laboratory \cite{NAP25171}, the Chinese Electron-Ion Collider of China \cite{chen2018plan}, or upgraded versions of CERN's LHC experiments, it is important to prepare adequate theoretical tools to compare and correctly interpret experimental results. One of the model frameworks allowing to estimate hadron structure functions is the combination of the McLerran-Venugopalan initial condition model together with the JIMWLK equation which describes the evolution in rapidity of the initial distribution. In this package we present a parallel C++ implementation of both these ingredients. In order to allow a thorough assessment of systematic effects several discretizations of the JIMWLK kernel are implemented both in position and momentum spaces. The effects of the running coupling in three different definitions are provided. The main code is supplemented with test and check programs for all main functionalities. The clear structure of the code allows easy implementation of further improvements such as the collinear constraint \cite{Hatta:2016ujq}. 
\end{abstract}


\end{frontmatter}

\section*{Code Metadata}
\label{sec. version}

\begin{center}
\begin{table}[H]
\begin{tabular}{|l|p{6.5cm}|p{6.5cm}|}
\hline
\textbf{Nr.} & \textbf{Code metadata description} &  \\
\hline
C1 & Current code version & 1.0 \\
\hline
C2 & Permanent link to code/repository used for this code version &  https://bitbucket.org/piotrekkorcyl/

jimwlk/src/master/\\
\hline
C3 & Code Ocean compute capsule & None \\
\hline
C4 & Legal Code License   & GNU general public license\\
\hline
C5 & Code versioning system used & git \\
\hline
C6 & Software code languages, tools, and services used & C++ \\
\hline
C7 & Compilation requirements, operating environments &  g++ compiler, openMPI, openMP, FFTW3 installations \\
\hline
C8 & Link to developer documentation/manual &  \\
\hline
C9 & Support email for questions & piotr.korcyl@uj.edu.pl\\
\hline
\end{tabular}
\caption{Code metadata}
\label{} 
\end{table}
\end{center}

\section{Motivation and Significance}
\label{sec. motivation}


Multiple experimental facilities designed to study the internal structure of hadrons are currently being planned. In particular, high-energy experiments such as the Electron-Ion Collider in the US, upgraded experiments at the LHC or Chinese Electron-Ion Collider of China promise extensive data about the three-dimensional, both in position and in momentum space, tomography of the quark and gluon composition of hadrons. From the theoretical point of view, the relevant observables can be obtained from several kinds of structure functions, ranging from the simplest parton distribution functions (PDF), through two-dimensional transverse momentum dependent (TMD) structure functions up to the most generic Wigner structure functions. In this work we limit ourselves to the TMD structure functions. For a given phenomenological process the latter can be obtained by combining together a class of distributions \cite{Bury:2018kvg} which can be evaluated numerically using the code provided in this package. 

The implemented calculations are based on the McLerran-Venugopalan (MV) model of the initial condition \cite{McLerran:1993ni,McLerran:1993ka}  coupled to an evolution equation in rapidity developed by Balitsky-Jalilian-Marian-Iancu-McLerran-Weigert-Leonidov-Kovner (JIMWLK) \cite{Balitsky:1995ub, JalilianMarian:1997jx,JalilianMarian:1997gr,JalilianMarian:1997dw,Kovner:2000pt,Kovner:1999bj,Weigert:2000gi,Iancu:2000hn,Ferreiro:2001qy}. Once the phenomenological parameters of the MV model are fixed, the framework allows to estimate all observables dependent on the TMD structure functions, in particular cross-sections. In view of the mentioned advances on the experimental side, it is important to accelerate progress on the theoretical side, which a common, open-source implementation of the MV model and the JIMWLK evolution equation should facilitate. Example of numerical results obtained with the discussed code package and their discussion is presented in Ref.~\cite{korcyl}.

The main object of studies are Wilson lines and their correlation functions. Wilson lines in the discretized setting are constructed from products of $N_y$ elementary Wilson links,
\begin{equation}
\label{eq. wilson line}
U^{ab}(\mathbf{x}) = \prod_{k=1}^{N_y} U^{ab}_k(\mathbf{x})
= \prod_{k=1}^{N_y} \exp\left( -i g A_k^{ab}(\mathbf{x}) \right)
    =\prod_{k=1}^{N_y} \exp\left(-i \frac{g \rho^{ab}_k(\mathbf{x})}{\nabla^2 + m^2} \right). 
\end{equation}
where the classical gauge potentials $A_k^{ab}$ are connected to the color sources $\rho^{ab}_k$ through a single Poisson equation,
\begin{equation}
    (\nabla^2 + m^2) g A^{ab}_k(\mathbf{x}) = g \rho^{ab}_k(\mathbf{x}).
\label{eq. poisson}
\end{equation}
which is explicitly imposed in Eq.~\eqref{eq. wilson line}.
The parameters $g$, $\mu$ belong to the MV model, whereas $N_y$, $m$ follow from a particular implementation of the involved physics. All dependence of final results on these quantities have been investigated in Ref.\cite{korcyl}. In its current state the code allows to evaluate the gluon distribution dependence on the transverse momentum at a given rapidity scale $s$ from the Wilson line correlation function
$C(\mathbf{x} - \mathbf{y}, s)$ \cite{Lappi:2007ku},
\begin{equation}
\label{eq. correlation function position}
C(\mathbf{x} - \mathbf{y},s) = \langle \textrm{tr}\, U^{\dagger}(\mathbf{x},s) U(\mathbf{y},s) \rangle,
\end{equation}
where the average $\langle \cdot \rangle$ is taken over
different statistical realizations of Wilson lines. The
evolution in rapidity is performed using the Langevin
formulation of the JIMWLK equation as explained in the
pionnering work by Weigert and Rummukainen
\cite{Rummukainen:2003ns} and Lappi \cite{Lappi:2012vw}. The
advance in rapidity $s$ by a step $\delta s$ is given by
\begin{multline}
\label{eq. main}
U(\mathbf{x},s+\delta s) = \exp\left( -\sqrt{\delta s} \sum_{\mathbf{y}} U(\mathbf{y},s) \left(\mathbf{K}(\mathbf{x}-\mathbf{y}) \cdot \boldsymbol{\xi}(\mathbf{y}) \right) U^{\dagger}(\mathbf{y},s) \right) \, \times \\ \times  U(\mathbf{x},s) \,  
\exp \left( \sqrt{\delta s} \sum_{\mathbf{y}} \mathbf{K}(\mathbf{x}-\mathbf{y}) \cdot \boldsymbol{\xi}(\mathbf{y}) \right),
\end{multline}
where $\mathbf{K}(\mathbf{x})$ is the JIMWLK kernel function and $\boldsymbol{\xi}(\mathbf{x})$ are random Gaussian vectors. 
\section{Overview of Capabilities}
\label{sec. capabilities}

The discretized problem is formulated on a transverse $x-y$ plane of size $L_x = L_y$, with a lattice spacing $a$. Although the evolution equation, given in Eq.~\eqref{eq. main}, is provided in position space, i.e. Wilson line at each position $\mathbf{x} = (x,y)$ of the transverse plain is evolved independently, the matrices appearing in the exponentials can be evaluated in position or momentum space. Since this requires JIMWLK kernels defined in both spaces, which can be discretized differently, the two formulations differ inherently by systematic effects which should be estimated and controlled. In the code package we provide both implementations.

It is known \cite{Rummukainen:2003ns} that for the results to be phenomenologically relevant the calculations have to include the effects of the running of the strong coupling constant. In the code we provide three different prescriptions how such effects can be taken into account. The physical discussion of the potential differences is presented in Ref.\cite{korcyl}.

\subsection{"Square root" prescription}

Following Rummukainen and Weingert \cite{Rummukainen:2003ns} we have introduced the running coupling with the coupling at the scale given by the size of the parent dipole $(x-y)^2$, i.e. we introduce a one-loop running
\begin{equation}
\label{eq. alpha_s}
    \alpha_s \rightarrow \alpha_s(1/(x-y)^2) = \frac{4 \pi}{\beta_0 \ln \frac{1}{(x-y)^2 \Lambda^2}}
\end{equation}
and we split the rapidity factor
$s = \frac{\alpha_s}{\pi^2}y$ with $y = \ln \frac{x_0}{x_2}$
by 
\begin{equation}
    \sqrt{\frac{\alpha_s \delta y}{\pi}} \rightarrow \sqrt{\frac{\delta y}{\pi}} \sqrt{\alpha_s(|x-y|)}
\end{equation}
The effects of the running coupling constant are accounted for in evolution in both, position and momentum spaces.

\subsection{Noise prescription}

The alternative definition of the running coupling was proposed by Lappi and Mantysaari \cite{Lappi:2012vw,Lappi:2014wya}. The running coupling can be implemented as a
modification of the properties of the noise vectors in the Langevin equation. 
Hence, instead of uncorrelated noise vectors $\xi$ in Eq. \ref{eq. main} we use
\begin{equation}
\langle \eta_{\mathbf{x}}^{a,i} \eta_{\mathbf{y}}^{b,j} \rangle = \delta^{a,b} \delta^{i,j} \int \frac{d^2\mathbf{k}}{(2\pi)^2} e^{i \mathbf{k}(\mathbf{x} - \mathbf{y})} \alpha_s(\mathbf{k}) = \delta^{a,b} \delta^{i,j} \tilde{\alpha}_{\mathbf{x} - \mathbf{y}}.
\end{equation}
The noise becomes correlated in both, momentum and position, spaces, and we provide implementations in both spaces. In momentum space the construction of the $\eta$ vectors with required properties is rather straightforward, because 
\begin{equation}
\langle \eta_{\mathbf{p}}^{a,i} \eta_{\mathbf{q}}^{b,j} \rangle = \delta^{a,b} \delta^{i,j} \delta(\mathbf{q} - \mathbf{p}) \alpha_s(\mathbf{p}).
\end{equation}
Contrary to that, in position space the nontrivial correlations exist between any two lattice sites
\begin{equation}
\langle \eta_{\mathbf{x}}^{a,i} \eta_{\mathbf{y}}^{b,j} \rangle = \delta^{a,b} \delta^{i,j} \tilde{\alpha}_{\mathbf{x} - \mathbf{y}}.
\end{equation}
These can be implemented through a construction of the desired correlation matrix
\begin{equation}
\Sigma(\mathbf{x},\mathbf{y}) = \tilde{\alpha}_s(\mathbf{x}-\mathbf{y}) \equiv \int \frac{d^2\mathbf{k}}{(2\pi)^2} e^{i \mathbf{k}(\mathbf{x} - \mathbf{y})} \alpha_s(\mathbf{k}),
\label{eq. sigma}
\end{equation}
which is subsequently decomposed using the Cholesky decomposition into $L$ such that $L L^{T} = \Sigma$.
Eventually, the required noise vectors in position space are obtained from uncorrelated noise vectors by a linear transformation 
\begin{equation}
\eta(\mathbf{x}) = \sum_{\mathbf{y}} L(\mathbf{x},\mathbf{y}) \xi(\mathbf{y}).
\end{equation}

\subsection{Hatta-Iancu prescription}

Finally, following Ref.\cite{Hatta:2016ujq} we provide yet another prescription of the running coupling constant, this time only in position space. In this prescription for each distance $r$ in the correlation function of Eq.~\eqref{eq. correlation function position}  between the Wilson lines, we perform the evolution as in the `square root` definition, with the modification that the coupling constant is evaluated at the scale $\alpha_s = \alpha_s( \min\{ |\mathbf{x}-\mathbf{y}|, r \} ).$

\section{Software Description}
\label{sec. description}

The code operates on two fundamental classes: 
\begin{verbatim}
template<class T, int t>  class lfield{
        std::complex<T>* u;
        int Lxl, Lyl;
        };
\end{verbatim}
and
\begin{verbatim}
template<class T, int t>  class gfield{
        std::complex<T>* u;
        int Lx, Ly;
        };
\end{verbatim}
describing the $t$-component field of complex numbers of type $T$. The \emph{global} field is defined on the entire lattice with $L_x \times L_y$ sites, whereas the \emph{local} version corresponds to the piece of the global field operated on by a single MPI process. In the latter case we assume that the entire problem is solved using many processes and each direction, $L_x$ and $L_y$ was partitioned into a number of pieces. Wilson lines being SU$(3)$ matrices are instantiated as $9$-component fields whereas real-valued correlation functions as $1$-component fields. All operations needed for the construction of the initial condition, evolution equation and estimation of final correlation function are implemented as class member functions, or external functions operating on fields. In order to handle the correlation matrix $\Sigma$ Eq.~\eqref{eq. sigma}, two analog classes, \verb[lmatrix[ and \verb[gmatrix[ have been implemented. For the encapsulation sake, all remaining parameters (MV model, parallelization) and objects (MPI exchange functions, Fourier transforms) have been implemented as separate classes.

The code provides two \verb[main[ files which cover all the possibilities: momentum vs. position evolution and three prescriptions implementations of the running coupling.  Apart of the optimized implementation invoked from the \verb[main_optimized.cpp[ function, there exists an analogous implementation in \verb[main_explicit.cpp[ where for each conceptual step of the construction of the Wilson lines and for each quantity appearing in the evolution equation, a separate object or function is provided and intermediate results are stored in separate objects. Although, this way of writing the code clearly presents the physical quantities existing behind the instantiated classes, it is not optimal from the memory utilization and thread creation and synchronization points of view. In the optimized version thread creation is limited to the minimum which requires merging of many operations into a single block and hence a more cryptic code. Independently, most of the important functions have a separate program intended to test and check their correctness. All test programs are collected in the separate \verb[tests[ folder.

Simulation parameters can be set from a provided configuration file, where the different evolution implementations and coupling constant prescriptions can be selected from an \verb[enum[ type,
\begin{verbatim}
enum Evolution { 
    POSITION_EVOLUTION, 
    MOMENTUM_EVOLUTION };
enum Coupling { 
    SQRT_COUPLING_CONSTANT, 
    NOISE_COUPLING_CONSTANT, 
    HATTA_COUPLING_CONSTANT };
enum Kernel { 
    LINEAR_KERNEL, 
    SIN_KERNEL };
\end{verbatim}

\section{Illustrative Examples}
\label{sec. examples}

\begin{figure}
\begin{center}
\includegraphics[width=0.49\textwidth]{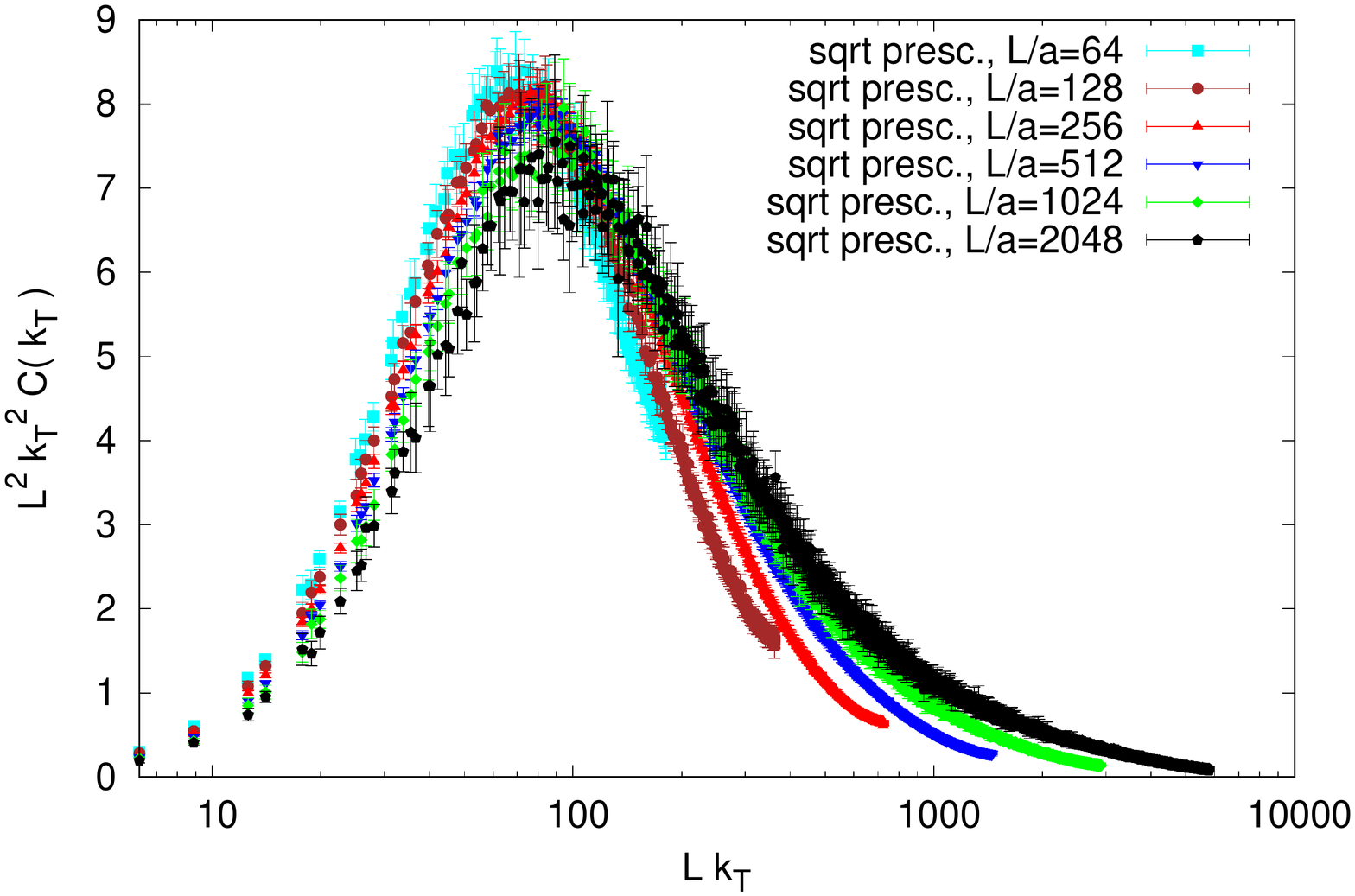}
\includegraphics[width=0.49\textwidth]{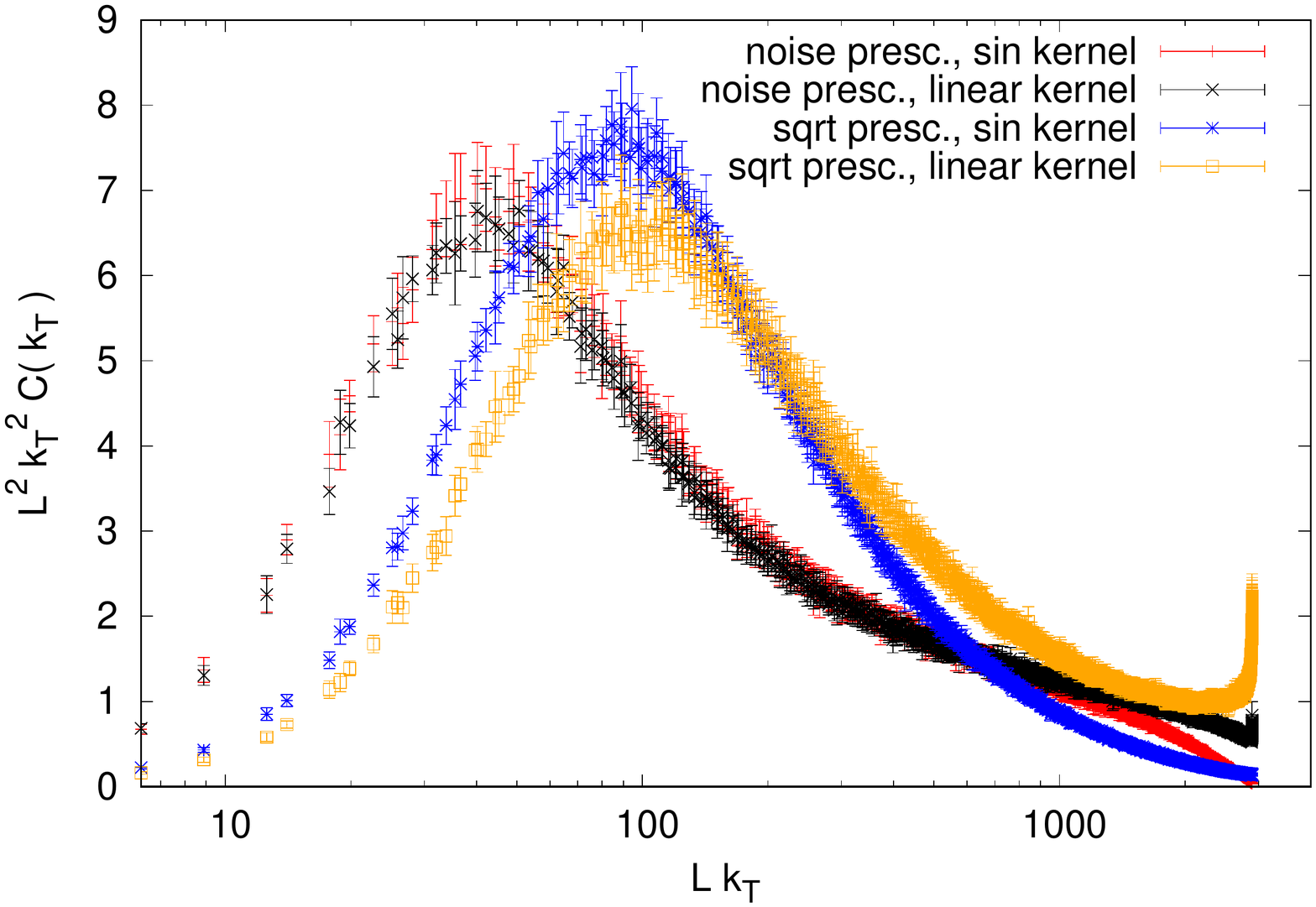}
\includegraphics[width=0.49\textwidth]{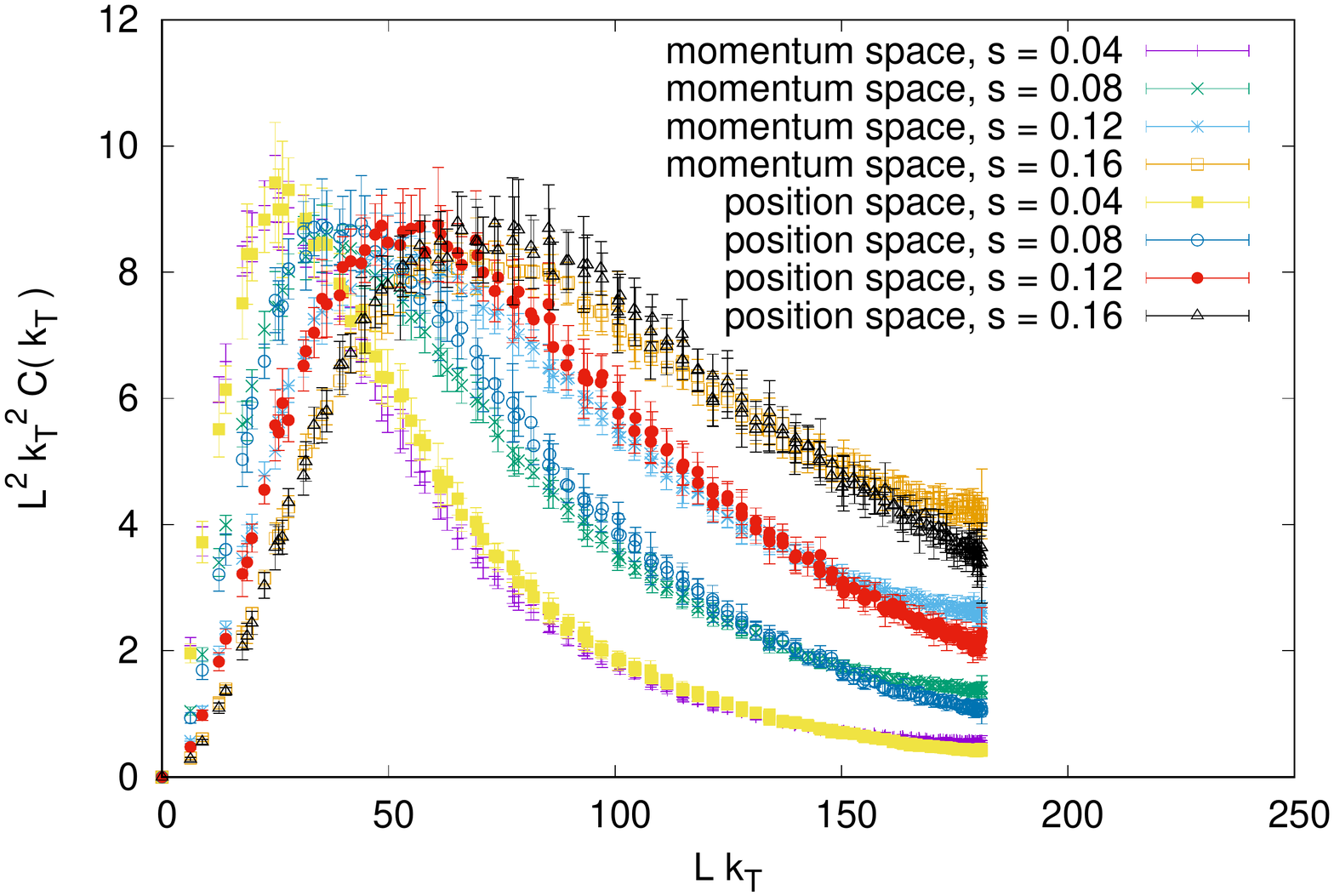}
\includegraphics[width=0.49\textwidth]{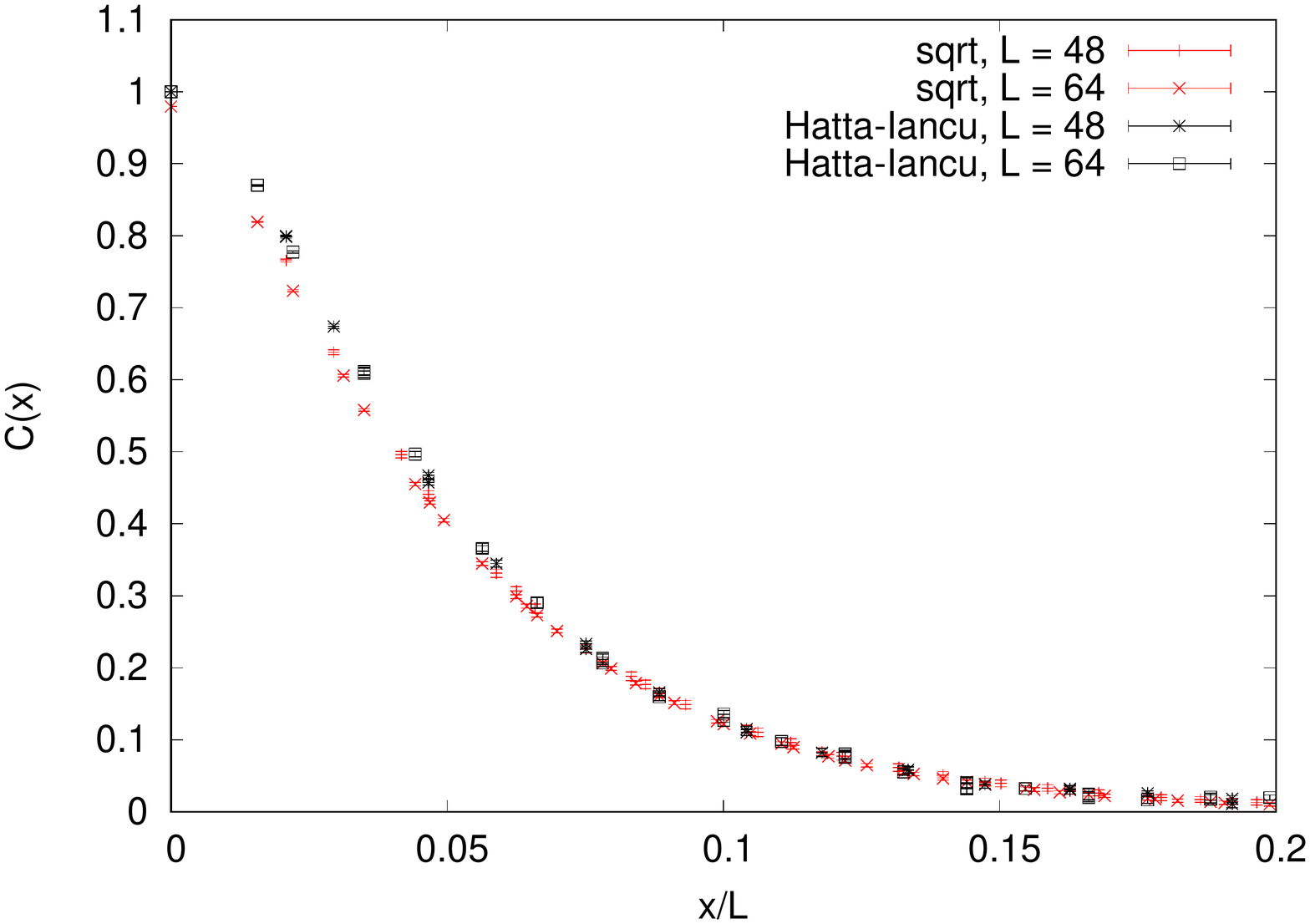}
\caption{Gluon distributions obtained by different discretizations and different implementations of the running coupling constant available in the code package. The value of $L k_T$ on the horizontal axis for which the distribution has its maximum can be used as an observable for the saturation scale $L Q_s$. The presented figures show dependencies of the numerical solutions of the JIMWLK equation on various algorithmic parameters: the dependence on the lattice extent (upper left panel), the dependence on the kernel discretization (upper right panel), the dependence on the position/momentum space implementation (lower left panel) and ''square root" and Hatta-Iancu running coupling constant prescription (lower right panel). The discussion of the physical aspects of these results are discussed in Ref.~\cite{korcyl}. \label{fig. example}}
\end{center}
\end{figure}

The simplest observable obtained with the package is the gluon distribution as a function of the transverse momentum. We provide an example of such distribution in Figure \ref{fig. example}, where the gluon distribution evolved to rapidity of $s=0.04$ is shown. The figure presents a comparison of the many different implementations included in the package and demonstrates the size of possible systematic effects associated with different algorithmic implementations. The thorough discussion of systematic effects and their physical interpretation can be found in Ref.~\cite{korcyl}.

\section{Performance}
\label{sec. performance}

\begin{figure}
\begin{center}
\includegraphics[width=0.95\textwidth]{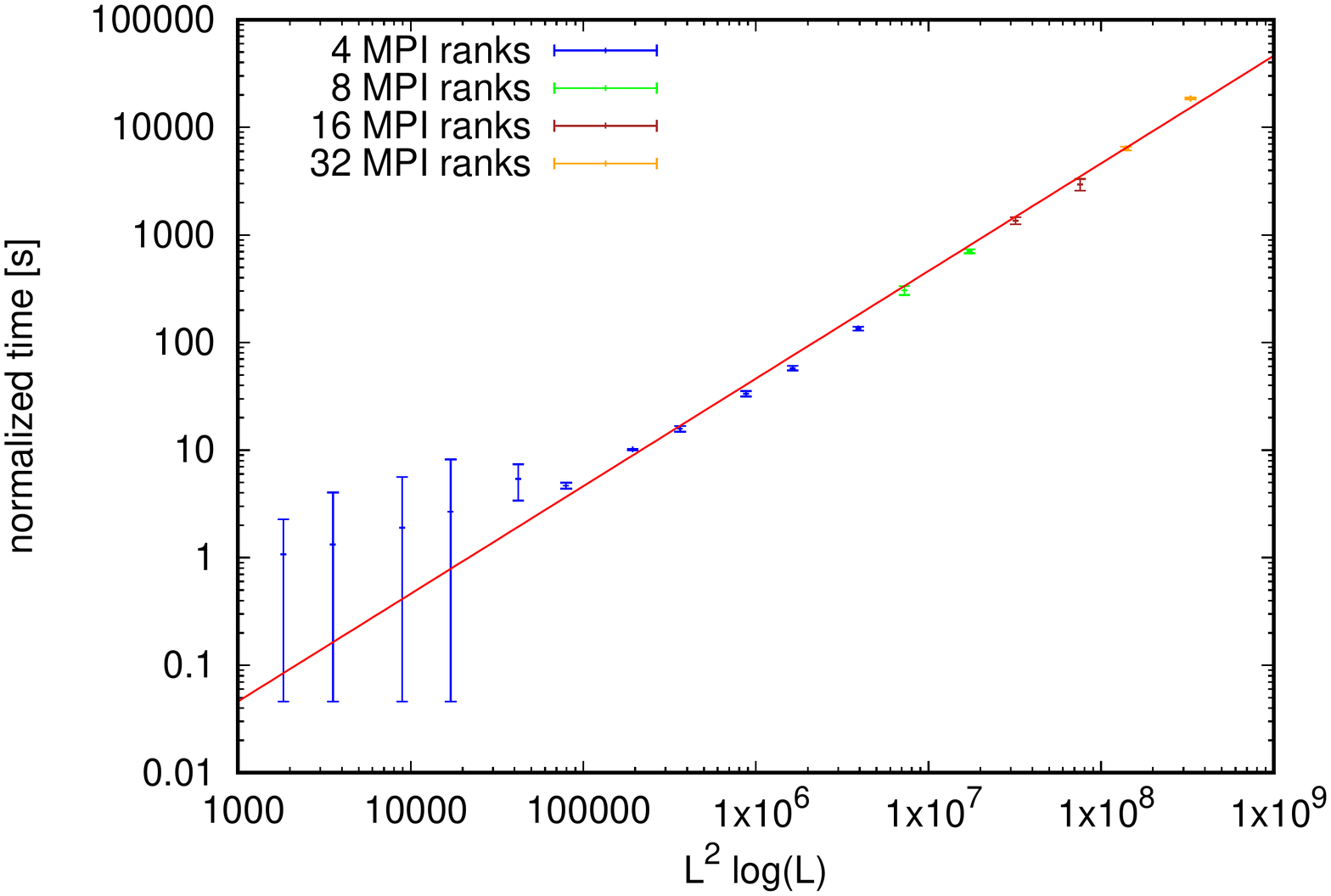}
\caption{Scaling with volume and parallelization. The times on the vertical axis correspond to the corresponding time on a single node. Volumes 24-768 were run on 4 nodes, 1024 and 1536 on 8 nodes, 2048 and 3072 on 16 nodes and 4086 and 6144 on 32 nodes. The uncertainties of the data points correspond to the minimal and maximal time within a single job. The large fluctuations for small volumes demonstrate the system jitter. \label{fig. scaling}}
\end{center}
\end{figure}

The code is prepared for High Performance Computing facilities. It is fully parallelized for both inter- and intra-node communications using MPI and openMP technology. The examplary scaling shown on Figure \ref{fig. scaling} was obtained on the Prometheus installation in CYFRONET Krak\'ow equipped with 24-cores Intel Haswell processors. The gathered timing corresponds to the evolution of the initial condition to the rapidity of $s=0.04$ using the implementation in momentum space. The numerical cost of the latter is saturated by multiple two-dimensional FFT transforms and hence scales as $L^2 \log(L)$. The position space implementation follows the expected $L^4$ scaling which was confirmed in numerical experiments.

\section{Impact}
\label{sec. impact}

Impact of this open-source package is expected to be three-fold.

From the phenomenological perspective the presented code offers a tool which can serve as a numerical model to estimate observables relevant for the planning of future high-energy experiments. The existing experimental data can be used to fit empirical parameters of the McLerran-Venugopalan model in a spirit recently explained in Ref. \cite{Mantysaari:2018zdd}. Afterwards this allows to generate gluon distributions at relevant values of the rapidity parameter and use them to evaluate expectations for appropriate cross-sections. The inclusion of multiple implementations of the evolution equation and especially multiple prescriptions for the running coupling effects allows for a thorough study of systematic uncertainties. The knowledge of the latter is crucial for predictions covering rapidity ranges not included in the initial data sets. Such theoretical calculations are necessary for the optimal planning of future experiments.

The open-source nature of the package facilitates cross-checks and enhances reproducibility of the results obtained by different groups. So far the numerical details of various independent implementations were either missing or scattered between different publications. This package offers a common, high-performance optimized basis, stored in a single \verb[git[ repository, which should provide a quick way for bug finding and fixing and would be beneficial for all involved groups.

Last but not least, the code in its current form is fully operational and allows calculations for the simplest two-point correlation function. The clean, highly-abstract implementation allows easy extensions. In particular, correlators of other operators constructed out of Wilson lines, such as the ones discussed in Ref.~\cite{Marquet:2016cgx} in the case of fixed coupling constant, can be added in a straightforward manner. Implementation of other, more complicated correlation functions should not pose problems, as the basis for MPI point-to-point boundary data exchanges is provided. This allows to attack new physical observables and hence provide new ways of connecting experimental data to theoretical calculations. We believe that further extensions of the entire framework, such as the ones proposed in Ref.~\cite{Cougoulic:2020tbc} or Ref.~\cite{Hatta:2016ujq}, could be collaboratively undertaken based on the presented code.

\section{Conclusions}
\label{sec. conclusions}

We have presented a C++ implementation of the MV model coupled to the JIMWLK evolution equations. The code allows to estimate transverse momentum dependent two-point gluon distributions. Other many-point correlation functions can be relatively easily supplemented. The precise knowledge of the latter is limited due to the insufficient coverage of experimental data in the rapidity variable. This fact, combined with the observation that the relevant physical regime of small-$x$ where non-perturbative Monte Carlo simulations are not feasible, hints that model calculations such as the ones proposed here, provide the sole theoretical framework allowing to produce predictions in the ranges of the envisaged high-energy facilities. It is therefore important to prepare well-tested, easily extensible numerical codes, which also allow modern, repository-based maintenance. We believe that the present package can be a first step into this direction.

\section*{Acknowledgements}
\label{sec. acknowledgements}
The Author is grateful for the help of K. Cichy at the early stages of the code implementation and T. Lappi for useful discussions. This work was in part supported by Deutsche  Forschungsgemeinschaft under Grant No.SFB/TRR 55 and by the polish NCN grant No. UMO-2016/21/B/ ST2/01492. The Author thanks for hospitality and fruitful discussions with C. Marquet and C. Roiesnel. This work was finalized during a visit at the Universita  degli  Studi  di  Roma  Tor  Vergata  which was made possible by the polish NAWA agency through the Bekker  fellowship. Numerical simulations were performed on the Prometheus supercomputer at CYFRONET AGH in Krak\'ow using the \emph{nspt}, \emph{pionda}, \emph{tmdlangevin} and \emph{plgtmdlangevin2} computer time allocations.

\bibliographystyle{elsarticle-num} 
\bibliography{ref}

\begin{thebibliography}{10}
\expandafter\ifx\csname url\endcsname\relax
  \def\url#1{\texttt{#1}}\fi
\expandafter\ifx\csname urlprefix\endcsname\relax\def\urlprefix{URL }\fi
\expandafter\ifx\csname href\endcsname\relax
  \def\href#1#2{#2} \def\path#1{#1}\fi

\bibitem{NAP25171}
{National Academies of Sciences, Engineering and Medicine},
  \href{https://www.nap.edu/catalog/25171/an-assessment-of-us\\-based-electron-ion-collider-science}{An
  Assessment of U.S.-Based Electron-Ion Collider Science}, The National
  Academies Press, Washington, DC, 2018.
\newblock \href {https://doi.org/10.17226/25171} {\path{doi:10.17226/25171}}.
\newline\urlprefix\url{https://www.nap.edu/catalog/25171/an-assessment-of-us\\-based-electron-ion-collider-science}

\bibitem{chen2018plan}
X.~Chen, A plan for electron ion collider in china (2018).
\newblock \href {http://arxiv.org/abs/1809.00448} {\path{arXiv:1809.00448}}.

\bibitem{Hatta:2016ujq}
Y.~Hatta, E.~Iancu, {Collinearly improved JIMWLK evolution in Langevin form},
  JHEP 08 (2016) 083.
\newblock \href {http://arxiv.org/abs/1606.03269} {\path{arXiv:1606.03269}},
  \href {https://doi.org/10.1007/JHEP08(2016)083}
  {\path{doi:10.1007/JHEP08(2016)083}}.

\bibitem{Bury:2018kvg}
M.~Bury, P.~Kotko, K.~Kutak,
  \href{http://dx.doi.org/10.1140/epjc/s10052-019-6652-4}{Tmd gluon
  distributions for multiparton processes}, The European Physical Journal C
  79~(2) (Feb 2019).
\newblock \href {https://doi.org/10.1140/epjc/s10052-019-6652-4}
  {\path{doi:10.1140/epjc/s10052-019-6652-4}}.
\newline\urlprefix\url{http://dx.doi.org/10.1140/epjc/s10052-019-6652-4}

\bibitem{McLerran:1993ni}
L.~D. McLerran, R.~Venugopalan, {Computing quark and gluon distribution
  functions for very large nuclei}, Phys. Rev. D49 (1994) 2233--2241.
\newblock \href {http://arxiv.org/abs/hep-ph/9309289}
  {\path{arXiv:hep-ph/9309289}}, \href
  {https://doi.org/10.1103/PhysRevD.49.2233}
  {\path{doi:10.1103/PhysRevD.49.2233}}.

\bibitem{McLerran:1993ka}
L.~D. McLerran, R.~Venugopalan, {Gluon distribution functions for very large
  nuclei at small transverse momentum}, Phys. Rev. D49 (1994) 3352--3355.
\newblock \href {http://arxiv.org/abs/hep-ph/9311205}
  {\path{arXiv:hep-ph/9311205}}, \href
  {https://doi.org/10.1103/PhysRevD.49.3352}
  {\path{doi:10.1103/PhysRevD.49.3352}}.

\bibitem{Balitsky:1995ub}
I.~Balitsky, {Operator expansion for high-energy scattering}, Nucl. Phys. B463
  (1996) 99--160.
\newblock \href {http://arxiv.org/abs/hep-ph/9509348}
  {\path{arXiv:hep-ph/9509348}}, \href
  {https://doi.org/10.1016/0550-3213(95)00638-9}
  {\path{doi:10.1016/0550-3213(95)00638-9}}.

\bibitem{JalilianMarian:1997jx}
J.~Jalilian-Marian, A.~Kovner, A.~Leonidov, H.~Weigert, {The BFKL equation from
  the Wilson renormalization group}, Nucl. Phys. B504 (1997) 415--431.
\newblock \href {http://arxiv.org/abs/hep-ph/9701284}
  {\path{arXiv:hep-ph/9701284}}, \href
  {https://doi.org/10.1016/S0550-3213(97)00440-9}
  {\path{doi:10.1016/S0550-3213(97)00440-9}}.

\bibitem{JalilianMarian:1997gr}
J.~Jalilian-Marian, A.~Kovner, A.~Leonidov, H.~Weigert, {The Wilson
  renormalization group for low x physics: Towards the high density regime},
  Phys. Rev. D59 (1998) 014014.
\newblock \href {http://arxiv.org/abs/hep-ph/9706377}
  {\path{arXiv:hep-ph/9706377}}, \href
  {https://doi.org/10.1103/PhysRevD.59.014014}
  {\path{doi:10.1103/PhysRevD.59.014014}}.

\bibitem{JalilianMarian:1997dw}
J.~Jalilian-Marian, A.~Kovner, H.~Weigert, {The Wilson renormalization group
  for low x physics: Gluon evolution at finite parton density}, Phys. Rev. D59
  (1998) 014015.
\newblock \href {http://arxiv.org/abs/hep-ph/9709432}
  {\path{arXiv:hep-ph/9709432}}, \href
  {https://doi.org/10.1103/PhysRevD.59.014015}
  {\path{doi:10.1103/PhysRevD.59.014015}}.

\bibitem{Kovner:2000pt}
A.~Kovner, J.~G. Milhano, H.~Weigert, {Relating different approaches to
  nonlinear QCD evolution at finite gluon density}, Phys. Rev. D62 (2000)
  114005.
\newblock \href {http://arxiv.org/abs/hep-ph/0004014}
  {\path{arXiv:hep-ph/0004014}}, \href
  {https://doi.org/10.1103/PhysRevD.62.114005}
  {\path{doi:10.1103/PhysRevD.62.114005}}.

\bibitem{Kovner:1999bj}
A.~Kovner, J.~G. Milhano, {Vector potential versus color charge density in low
  x evolution}, Phys. Rev. D61 (2000) 014012.
\newblock \href {http://arxiv.org/abs/hep-ph/9904420}
  {\path{arXiv:hep-ph/9904420}}, \href
  {https://doi.org/10.1103/PhysRevD.61.014012}
  {\path{doi:10.1103/PhysRevD.61.014012}}.

\bibitem{Weigert:2000gi}
H.~Weigert, {Unitarity at small Bjorken x}, Nucl. Phys. A703 (2002) 823--860.
\newblock \href {http://arxiv.org/abs/hep-ph/0004044}
  {\path{arXiv:hep-ph/0004044}}, \href
  {https://doi.org/10.1016/S0375-9474(01)01668-2}
  {\path{doi:10.1016/S0375-9474(01)01668-2}}.

\bibitem{Iancu:2000hn}
E.~Iancu, A.~Leonidov, L.~D. McLerran, {Nonlinear gluon evolution in the color
  glass condensate. 1.}, Nucl. Phys. A692 (2001) 583--645.
\newblock \href {http://arxiv.org/abs/hep-ph/0011241}
  {\path{arXiv:hep-ph/0011241}}, \href
  {https://doi.org/10.1016/S0375-9474(01)00642-X}
  {\path{doi:10.1016/S0375-9474(01)00642-X}}.

\bibitem{Ferreiro:2001qy}
E.~Ferreiro, E.~Iancu, A.~Leonidov, L.~McLerran, {Nonlinear gluon evolution in
  the color glass condensate. 2.}, Nucl. Phys. A703 (2002) 489--538.
\newblock \href {http://arxiv.org/abs/hep-ph/0109115}
  {\path{arXiv:hep-ph/0109115}}, \href
  {https://doi.org/10.1016/S0375-9474(01)01329-X}
  {\path{doi:10.1016/S0375-9474(01)01329-X}}.

\bibitem{korcyl}
S.~Cali, K.~Cichy, P.~Korcyl, P.~Kotko, K.~Kutak, C.~Marquet, On systematic
  effects in the numerical solutions of the jimwlk equation (2021).
\newblock \href {http://arxiv.org/abs/2104.14254} {\path{arXiv:2104.14254}}.

\bibitem{Lappi:2007ku}
T.~Lappi, {Wilson line correlator in the MV model: Relating the glasma to deep
  inelastic scattering}, Eur. Phys. J. C55 (2008) 285--292.
\newblock \href {http://arxiv.org/abs/0711.3039} {\path{arXiv:0711.3039}},
  \href {https://doi.org/10.1140/epjc/s10052-008-0588-4}
  {\path{doi:10.1140/epjc/s10052-008-0588-4}}.

\bibitem{Rummukainen:2003ns}
K.~Rummukainen, H.~Weigert, {Universal features of JIMWLK and BK evolution at
  small x}, Nucl. Phys. A739 (2004) 183--226.
\newblock \href {http://arxiv.org/abs/hep-ph/0309306}
  {\path{arXiv:hep-ph/0309306}}, \href
  {https://doi.org/10.1016/j.nuclphysa.2004.03.219}
  {\path{doi:10.1016/j.nuclphysa.2004.03.219}}.

\bibitem{Lappi:2012vw}
T.~Lappi, H.~Mäntysaari, {On the running coupling in the JIMWLK equation},
  Eur. Phys. J. C73~(2) (2013) 2307.
\newblock \href {http://arxiv.org/abs/1212.4825} {\path{arXiv:1212.4825}},
  \href {https://doi.org/10.1140/epjc/s10052-013-2307-z}
  {\path{doi:10.1140/epjc/s10052-013-2307-z}}.

\bibitem{Lappi:2014wya}
T.~Lappi, H.~Mäntysaari, {Proposal for a running coupling JIMWLK equation},
  Nucl. Phys. A932 (2014) 69--74.
\newblock \href {http://arxiv.org/abs/1403.7289} {\path{arXiv:1403.7289}},
  \href {https://doi.org/10.1016/j.nuclphysa.2014.07.009}
  {\path{doi:10.1016/j.nuclphysa.2014.07.009}}.

\bibitem{Mantysaari:2018zdd}
H.~Mäntysaari, B.~Schenke, {Confronting impact parameter dependent JIMWLK
  evolution with HERA data}, Phys. Rev. D 98~(3) (2018) 034013.
\newblock \href {http://arxiv.org/abs/1806.06783} {\path{arXiv:1806.06783}},
  \href {https://doi.org/10.1103/PhysRevD.98.034013}
  {\path{doi:10.1103/PhysRevD.98.034013}}.

\bibitem{Marquet:2016cgx}
C.~Marquet, E.~Petreska, C.~Roiesnel, {Transverse-momentum-dependent gluon
  distributions from JIMWLK evolution}, JHEP 10 (2016) 065.
\newblock \href {http://arxiv.org/abs/1608.02577} {\path{arXiv:1608.02577}},
  \href {https://doi.org/10.1007/JHEP10(2016)065}
  {\path{doi:10.1007/JHEP10(2016)065}}.

\bibitem{Cougoulic:2020tbc}
F.~Cougoulic, Y.~V. Kovchegov, {Helicity-dependent extension of the
  McLerran-Venugopalan model} (5 2020).
\newblock \href {http://arxiv.org/abs/2005.14688} {\path{arXiv:2005.14688}}.

\end{thebibliography}

\end{document}